\documentstyle[axodraw]{article}
\def\1ad{\mbox{\normalsize $^1$}}
\def\2ad{\mbox{\normalsize $^2$}}
\def\3ad{\mbox{\normalsize $^3$}}
\def\4ad{\mbox{\normalsize $^4$}}
\def\5ad{\mbox{\normalsize $^5$}}
\def\6ad{\mbox{\normalsize $^6$}}
\def\7ad{\mbox{\normalsize $^7$}}
\def\8ad{\mbox{\normalsize $^8$}}
\def\makefront{
\begin{flushright}
\small
AEI-104, SPIN-1999/04\\
hep-th/9903099
\end{flushright}
\vskip  0.2truecm
\begin{center}
\def\newtitleline{\\ \vskip 5pt}
{\Large\bf\titleline}\\
\vskip 1truecm
{\large\bf\authors}\\
\vskip 5truemm
\addresses
\end{center}
\vskip 1truecm
{\bf Abstract:}
\abstracttext
\vskip 1truecm}
\def\bea{\begin{eqnarray}}
\def\eea{\end{eqnarray}}
\def\be{\begin{equation}}
\def\ee{\end{equation}}

\newcommand\rf[1]{(\ref{#1})}
\def\nn{\nonumber}

\newcommand{\NPB}[3]{{Nucl.\ Phys.} {\bf B#1} (#2) #3}

\newcommand{\PRD}[3]{{Phys.\ Rev.} {\bf D#1} (#2) #3}
\newcommand{\PRL}[3]{{Phys.\ Rev.\ Lett.} {\bf #1} (#2) #3}

\newcommand{\JHEP}[3]{{JHEP} {\bf #1} (#2) #3}
\newcommand{\ft}[2]{{\textstyle\frac{#1}{#2}}\,}

\newcommand{\svv}{{v_{12}}\!\!\!\!\!\!\! /\,\,\,\,\,}

\newcommand{\sq}{{q_1}\!\!\!\!\! /\,\,\,}
\newcommand{\tr}{\mbox{tr}}

\begin {document}
\def\titleline{
Matrix Theory and Feynman Diagrams\footnote{Talk given by J. Plefka
at the 32nd Symposium Ahrenshoop on the Theory of Elementary Particles
Buckow, September 1 - 5, 1998.}
}
\def\authors{
J. Plefka\1ad, M. Serone\2ad and A. Waldron\3ad
}
\def\addresses{
\1ad Albert-Einstein-Institut\\
Max-Planck-Institut f\"ur Gravitationsphysik\\
Schlaatzweg 1, 14473 Potsdam,
Germany\\ 
\2ad Department of Mathematics, University of Amsterdam\\
Plantage Muidergracht 24, 1018 TV Amsterdam\\
The Netherlands\\ 
\3ad NIKHEF, P.O. Box 41882, 1009 DB Amsterdam\\
The Netherlands\\
}
\def\abstracttext{
We briefly review the computation of graviton and antisymmetric
tensor scattering amplitudes
in Matrix Theory from a diagramatic S-Matrix point of view.
}
\large
\makefront
It is by now commonly believed that eleven dimensional supergravity
is the low-energy effective theory of a more fundamental microscopic
model, known as M-theory. A proposed definition of M-theory
has been given in terms of the large $N$ limit of  
a quantum mechanical supersymmetric $U(N)$ Yang-Mills model called
Matrix
Theory \cite{bfss}. Since the time of this conjecture many computations
have been performed to test the proposal \cite{MT}. 
We here wish to focus on the particle like excitations of Matrix Theory, 
namely the 11-dim supergravity
multiplet, and review a setup which enables one to compute various
S-Matrix
elements amongst them \cite{us1,us2}. 
These amplitudes are then shown to
agree with the corresponding tree-level amplitudes of the low energy 
supergravity theory, thus demonstrating that Matrix Theory indeed
has D=11 supergravity as its low energy limit. We emphasize that
it is possible to perform this comparison on the familiar level of
Feynman diagram amplitudes in momentum space, in particular there
is no need to revert to the calculation of effective source-probe
potentials on the supergravity side once one employs our definition
of the Matrix Theory S-Matrix. This approach appears to be mandatory
once one intends to go beyond the tree-level (classical) approximation
to
study quantum corrections to the effective M-theory action. After
all this is the true challenge for Matrix Theory, as the four
and six point tree-level amplitudes are believed to be determined
by the large amount of supersymmetry in the problem \cite{Sethi}.

\section{The Matrix Theory S-Matrix}

The M-theory matrix model emerges as the reduction of 9+1 dimensional
supersymmetric $U(N)$ Yang-Mills to 0+1 dimensions. In euclidean
time the quantum mechanical action reads
\be
S=\int dt \Bigl [ \frac 1 2 \tr (D_t X^a)^2 - \frac 1 4 \tr [X^a,X^b]^2
+ \frac 1 2 \tr ( \psi^T D_t \psi - \psi^T \gamma_a [X^a,\psi] ) \Bigr
],
\ee
where $D_t X^a=\partial_tX^a-i[A,X^a]$ and $D_t
\psi=\partial_t\psi-i[A,\psi]$,
$A$, $X^a$ and $\psi_\alpha$ are hermitian $U(N)$ matrices,
($a=1,\ldots,9$
and $\alpha=1,\ldots, 16$). Moreover we employ a real symmetric 
representation 
for the Dirac matrices $\gamma_a$ in which the charge conjugation matrix 
$\cal{C}$ equals unity. One
proceeds to compute the background field effective action by expanding
the matrices $X^a_{ij}$ and $\psi_{ij}$ ($i=1,\ldots,N$) around the 
diagonal background
\be
X^a_{ij}=\delta_{ij}(v_i^a\, t + b_i^a) + Y^a_{ij} \qquad
\psi_{ij}=\delta_{ij}\theta_i + \tilde\psi_{ij} 
\ee
with the constant velocities $v_i$, impact paramters $b_i$ and
spins $\theta_i$. This background manifestly solves the classical
equations of motion. After fixing a standard background field
gauge the quantum fields $Y^a$, $\tilde\psi$ and $A$, 
as well as the 
ghost matrices $c$ and $\bar c$ are then integrated out 
of the path integral via a perturbative loop expansion. One finally
arrives 
at the effective background field action taking the form
\be
\int_{-T/2}^{T/2} dt\, \Gamma=\frac T 2 \sum_i v_i^2 + 
\int_{-T/2}^{T/2} dt \sum_{l\geq 1}
\Gamma^{(l)}(v_{ij},r_{ij}(t),\theta_{ij})
\ee
Note that in the $l$ loop effective potential $\Gamma^{(l)}$ only 
the relative quantities
$v_{ij}=v_i-v_j$, $r_{ij}(t)=v_{ij}t + b_{ij}$ and 
$\theta_{ij}=\theta_i-\theta_j$ enter, $T$ is the interaction time.

Let us from now on concentrate on the $U(2)$ model being relevant to two 
body interactions. At one loop the effective potential reads
$\Gamma^{(1)}=v_{12}^4/r_{12}(t)^7 + (\theta_{12}\mbox{ -terms})$.
In order to compute the corresponding
4-point S-Matrix of Matrix Theory, we prepare two incoming and two 
outgoing states
carrying the quantum numbers momentum $p$ and polarization $\cal{H}$
of the four particles involved in the scattering process. The S-Matrix
element is then written as
\be
{\cal S}_{2 \rightarrow 2}= \langle {{\cal H}^1}' , {{\cal H}^2}'|
\int (\prod_{i=1}^2 d^9(x'_i,x_i) e^{-i p_i\cdot x_i + i p'_i\cdot x'_i}
\exp(i\frac T 2 (v_1^2+v_2^2)+i\int_{-T/2}^{T/2}dt\, \Gamma^{(1)})
\,  |{\cal H}^1 , {\cal H}^2\rangle
\label{S-Matrix}
\ee
We will return shortly to the precise nature of the polarization
states $|\cal{H}\rangle$. In the above integral one now changes
variables
to $v_i=(x_i'-x_i)/T$ and $b_i=(x'_i+x_i)/2$ ($i=1,2$), and performs the
emerging $v_i$ integrals via stationary phase in the relevant 
$T\rightarrow \infty$ limit. This yields the relation
\be
v_i^a=\frac{{p^a_i}'+p^a_i}{2}\, .
\ee
Expanding out the exponential $\exp(\int dt \Gamma^{(1)})$ and
performing the
$t$ and $b_i$ integrals finally leads to the matrix element
\bea
\lefteqn{{\cal S}_{2 \rightarrow 2}= 
\delta^9(q_1^a+q_2^a)\,\delta(q_1\cdot v_1+ 
q_2\cdot v_2)\,\times
 {}_{\theta_1}\langle {{\cal H}^1}'|\, {}_{\theta_2}\langle {{\cal
H}^2}'|
\Bigl [\, v_{12}^4 + 2\,v_{12}^2 
(\theta_{12}\svv\sq\theta_{12})} \label{s1}\\&&
+2(\theta_{12}\svv\sq\theta_{12})^2
+ \frac{4}{9} (\theta_{12}\svv\sq\theta_{12})
(\theta_{12}\sq\gamma^{a}\theta_{12})^2
+ \frac{2}{63}\Bigl((\theta_{12}\sq\gamma^{a}\theta_{12})^2\Bigr )^2
\Bigl ]\, \frac{1}{q_1^2} \,  |{\cal H}^1\rangle_{\theta_1} {\cal H}^2
\rangle_{\theta_2}\nn
\eea
with the momentum transfer $q_i^a={p^a_i}'-p^a_i$. 
Here we have inserted the complete form 
of the $U(2)$ effective potential, including all spin dependent
terms \cite{it,f2}.
Note the emergence of 
the momentum and energy conserving $\delta$-functions in \rf{s1}. In
\rf{s1} the $\theta_{12}=\theta_1-\theta_2$ are 
to be interpreted as operators,
obeying the algebra
$\{\theta_i^\alpha,\theta_j^\beta\}=\delta^{\alpha\beta}
\delta_{ij}$. The polarization
states form a representation of this algebra representing the 256 states
of the d=11 supergraviton multiplet, i.e.
\be
|{\cal{H}}^1\rangle = {\cal{H}}^1_{{\cal M}} |-\rangle^{{\cal
M}}_{\theta_1},
\qquad
|{\cal{H}}^2\rangle = {\cal{H}}^2_{{\cal M}} |-\rangle^{{\cal
M}}_{\theta_2}
\ee
where the generalized index ${\cal M}$ denotes 
${\cal M}\equiv \{ab;abc;ab\alpha
\}$, corresponding to the graviton, three-form and gravitino,
respectively
\cite{QM}.
These states are built from the diagonal entries $\theta_1$ and
$\theta_2$,
contrary to the effective potential in \rf{s1} depending only on
$\theta_1
-\theta_2$. However, via a Fierz rearrangement we may always convert
to fermion bilinears homogenous in $\theta_1$ or $\theta_2$,
as long as we restrict our attention to amplitudes involving bosons
only.
The general case is a little bit more involved. Whereas 
$\theta_1\gamma^{ab}\theta_1$ just acts as a $SO(9)$ rotation on 
$|-\rangle^{ab}_{\theta_1}$ and $|-\rangle^{abc}_{\theta_1}$, the three
index operator acts as
\bea
\theta_1\gamma^{mnp}\theta_1|-\rangle^{tu}_{\theta_1}&=&
-i24\sqrt 3 (\delta^{mt}\,|-\rangle^{npu}_{\theta_1}
-\ft{1}{9}\delta^{tu}\,|-\rangle^{mnp}_{\theta_1})\,\, ,\label{alg1}\\
\theta_1\gamma^{mnp}\theta_1|-\rangle^{tuv}_{\theta_1}&=&
i24\sqrt 3\, \delta^{mt}\delta^{nu}\,|-\rangle^{pv}_{\theta_1}+
\ft{2i}{3}\epsilon^{mnptuvwyz}
|-\rangle^{wyz}_{\theta_1}\, .\label{alg2}
\eea
On the right hand side of \rf{alg1} and \rf{alg2} one must
(anti)symmetrize (with unit weight) over all indices according to
the symmetry  properties of the left hand sides of these equations.
Given the algebra of \rf{alg1} and \rf{alg2} only a moderate amount
of computer algebra is now required to obtain any pure bosonic S-matrix
element. We have computed the 4-graviton amplitude \cite{us1} consisting
of 66 terms as well as the 4-three form amplitude \cite{us2} made out of 
103 terms. Due to lack of space we do not state the resulting amplitudes
here.

\section{The Supergravity Computation}

The bosonic sector of 11-dimensional supergravity \cite{Cremmer} is  
given by
\bea
\lefteqn{{\cal L}= -\ft{1}{2\kappa^2} \sqrt{-g}\, R -\ft 18 \sqrt{-g}\,
(F_{MNPQ})^2}
\nn\\&& -\ft{\sqrt{3}}{12^3\kappa}\varepsilon^{M_1\ldots M_{11}}
F_{M_1M_2M_3M_4}\, F_{M_5 M_6M_7M_8}\, C_{M_9M_{10}M_{11}}
\label{sugra}
\eea
where $F_{MNPQ}=4\partial_{[M}C_{NPQ]}$,
$g={\rm det}\,g_{MN}$ and $M=0,\ldots,10$.
Perturbative
quantum gravity is studied by considering small fluctuations $h_{MN}$
from the flat metric $\eta_{MN}$, i.e. 
$
g_{MN}= \eta_{MN}+ \kappa\, h_{MN}
$
where $\kappa$ is the 11-dimensional gravitational coupling
constant. After employing (for example) the harmonic gauge 
$\partial_N h^N{}_M-(1/2)\partial_M h^N{}_N=0$ for the graviton as well
as the gauge $\partial_M C^M{}_{NP}=0$ for the antisymmetric 
tensor one derives the bosonic propagators in a straightforward
fashion. The relevant Feynman diagrams for the two amplitudes considered
are the $t=-2\, p^1\cdot{p^1}'$ channel ones, which dominate eikonal
physics. They are:
\begin{center}
\begin{picture}(320,60)(-20,-20)
\Gluon(0,30)(50,30){2}{6}
\Gluon(0,0)(50,0){2}{6}
\Gluon(25,30)(25,0){2}{3}
\Text(25,-15)[]{4-$g_{MN}$}
\Text(-10,30)[]{1}
\Text(60,30)[]{1'}
\Text(-10,0)[]{2}
\Text(60,0)[]{2'}

\Line(120,30)(170,30)
\Line(120,0)(170,0)
\Gluon(145,30)(145,0){2}{3}
\Text(110,30)[]{1}
\Text(180,30)[]{1'}
\Text(110,0)[]{2}
\Text(180,0)[]{2'}

\Line(250,30)(300,30)
\Line(250,0)(300,0)
\Line(275,30)(275,0)
\Text(240,30)[]{1}
\Text(310,30)[]{1'}
\Text(240,0)[]{2}
\Text(310,0)[]{2'}

\Text(210,-15)[]{4-$B_{MNP}$}
\end{picture}
\end{center}
In order to make a comparison with the Matrix Theory results
we need to rewrite our supergravity $t$ channel amplitudes
in terms of physical transverse nine dimensional degrees of freedom.
Since we are considering the $N=2$ discrete light cone
quantisation formulation~\cite{suss}
of the theory, we work in
light cone coordinates and specialise to the case of vanishing
$p^-$ momentum exchange. The kinematics of our 
$1+2\longrightarrow 1'+2'$ 
process then reads
\bea
p^M_1=(-\ft12\,(v_1^a-q_1^a/2)^2\, ,\, 1\, ,\, v_1^a-q_1^a/2)
&\!\!\!\!\!\!\!\!\!&
p^M_{1'}=(-\ft12\,(v_1^a+q_1^a/2)^2\, ,\, 1\, ,\, v_1^a+q_1^a/2)
\nonumber\\
p^M_2=(-\ft12\,(v_2^a-q_2^a/2)^2\, ,\, 1\, ,\, v_2^a-q_2^a/2) &\!\!\!\!
\!\!\!\!\!\!&
p^M_{2'}=(-\ft12\, (v_2^a+q_2^a/2)^2\, ,\, 1\, ,\,v_2^a+q_2^a/2)\nn\\&&
\label{kinematics}
\eea
in units of the compactified radius $R$ and where the 11-dim index $M$ 
decomposes as $M=(+,-,a)$. We then reduce to physical
polarizations by setting
$
h_{+M}=0=h^a_a
$, $h_{-M}=-p^a\, h_{aM}$ as well as $C_{+-a}=0=C_{+ab}$ and
$C_{-ab}=-\frac{1}{p^-}p_{c}C_{cab}$.
As an example the resulting four graviton amplitude up to terms 
of order $v_{12}^2$ reads 
\begin{eqnarray}
{\cal A}&=& \frac{\textstyle 1}{\textstyle {q_1}^2}\Biggl\{ 
\ft12(h_1 h_1')(h_2 h_2') v_{12}^4 
+ 2\Bigr[({q_1} h_2' h_2 v_{12}) (h_1 h_1') 
      - ({q_1} h_2 h_2' v_{12}) (h_1 h_1')\Bigr] v_{12}^2 
\nn\\&&
+  (v_{12}h_2v_{12}) ({q_1}h_2'{q_1})(h_1 h_1') 
+  (v_{12}h_2'v_{12}) ({q_1}h_2{q_1})(h_1 h_1') 
- 2({q_1}h_2v_{12}) ({q_1}h_2'v_{12})(h_1 h_1') 
\nn\\&&
- 2 ({q_1}h_1h_1'v_{12}) ({q_1}h_2'h_2v_{12})
+ ({q_1}h_1h_1'v_{12}) ({q_1}h_2h_2'v_{12}) 
+ ({q_1}h_1'h_1v_{12}) ({q_1}h_2'h_2v_{12})  
\nn\\&&
+ \ft{1}{2}\Bigl [({q_1}h_1h_1'h_2'h_2{q_1})
-  2({q_1}h_1h_1'h_2h_2'{q_1}) 
+  ({q_1}h_1'h_1h_2h_2'{q_1}) 
-  2({q_1}h_2h_2'{q_1})(h_1 h_1') \Bigr ] v_{12}^2 
\nn\\&&+  \ldots
\, \Biggr\}
+ \Bigl[h_1 \longleftrightarrow h_2\, , \, h_2' \longleftrightarrow h_1'
\Bigr]
\label{Ulle}
\end{eqnarray}
and agrees with the Matrix Theory result. The same holds true for
the four three-form amplitude. 

\end{document}